\documentclass[prb,preprint,superscriptaddress,floatfix]{revtex4-2}

\usepackage{amsmath}
\usepackage{bm}
\usepackage{tabularx}
\usepackage[dvipdfmx]{graphicx}
\usepackage{comment}
\usepackage{lineno}
\usepackage{dcolumn}
\usepackage{float}
\usepackage{enumerate}
\usepackage{multirow}
\usepackage{color}
\usepackage{comment}

\graphicspath{{Fig/}}

\begin{document}
\newcommand{\tn}{$T_{\rm N}$}
\newcommand{\tord}{$T_{\rm o}$}
\newcommand{\mub}{${\mu}_{B}$}
\newcommand{\upg}{U$_{2}$Pt$_{6}$Ga$_{15}$}
\newcommand{\rpx}{$R_{2}$Pt$_{6}X_{15}$}

\sloppy
\title{Magnetic Order in Honeycomb Layered \upg \ Studied by Resonant X-ray and Neutron Scatterings}

\author{Chihiro Tabata}
\affiliation{Materials Sciences Research Center, Japan Atomic Energy Agency, Tokai 319-1195, Japan}
\affiliation{Advanced Science Research Center, Japan Atomic Energy Agency, Tokai 319-1195, Japan}

\author{Fusako Kon} 
\affiliation{Graduate School of Science, Hokkaido University, Sapporo 060-0810, Japan}

\author{Kyugo Ota}
\affiliation{Graduate School of Science and Engineering for Education, University of Toyama, Toyama 930-8555, Japan}

\author{Ruo Hibino}
\affiliation{Graduate School of Science, Hokkaido University, Sapporo 060-0810, Japan}

\author{Yuji Matsumoto}
\affiliation{Graduate School of Science and Engineering, University of Toyama, Toyama 930-8555, Japan}

\author{Hiroshi Amitsuka}
\affiliation{Graduate School of Science, Hokkaido University, Sapporo 060-0810, Japan}

\author{Hironori Nakao}
\affiliation{Photon Factory, Institute of Materials Structure Science, High Energy Accelerator Organization, Tsukuba 305-0801, Japan}
\affiliation{Condensed Matter Research Center, Institute of Materials Structure Science, High Energy Accelerator Organization, Tsukuba 305-0801, Japan}

\author{Yoshinori Haga}
\affiliation{Advanced Science Research Center, Japan Atomic Energy Agency, Tokai 319-1195, Japan}

\author{Koji Kaneko}
\affiliation{Materials Sciences Research Center, Japan Atomic Energy Agency, Tokai 319-1195, Japan}
\affiliation{Advanced Science Research Center, Japan Atomic Energy Agency, Tokai 319-1195, Japan}

\date{\today}

\begin{abstract}
Antiferromagnetic (AF) order of {\upg} with the ordering temperature $T_{\rm N}$~=~26~K was investigated by resonant X-ray scattering and neutron diffraction on single crystals. 
This compound possesses a unique crystal structure in which uranium ions form honeycomb layers and then stacks along the $c$-axis with slight offset, which gives rise to a stacking disorder.
The AF order can be described with the propagation vector of $\bm{q} = (1/6, 1/6, 0)$ in the hexagonal notation. 
The ordered magnetic moments orient perpendicular to the honeycomb layers, indicating a collinear spin structure consistent with Ising-like anisotropy. 
The magnetic reflections are found to be broadened along $c^*$ indicating that the stacking disorder results in anisotropic correlation lengths.
The semi-quantitative analysis of neutron diffraction intensity, combined with group theory considerations based on the crystallographic symmetry, suggests a zig-zag type magnetic structure for the AF ground state,
in which the AF coupling runs perpendicular to the stacking offset, characterized as $\bm{q} = (1, 0, 0)_{\rm orth}$.
The realization of the zig-zag magnetic structure implies the presence of frustrating intralayer exchange interactions involving both ferromagnetic (FM) first-neighbor and AF second and third-neighbor interactions in this compound.
\end{abstract}

\maketitle

\section{Introduction}
Frustrated magnets have been a promising research field to accommodate diverse and intriguing phenomena, including topological magnetic order \cite{Nagaosa2013}, spin-liquid states \cite{Balents2010}, quantum Hall effect \cite{Peng1988,Ueland2012}, and unconventional superconductivity \cite{Glasbrenner2015,Watanabe2019}.
When only the nearest neighbor antiferromagnetic (AF) interaction, $J_1$, is effective, magnetic frustration arises in triangular lattice and their derivatives such as Kagom\'{e} and pyrochlore lattices. 
In contrast, a honeycomb network does not have such frustrations since an AF $J_1$ interaction results in a N\'{e}el type AF order.
However, magnetic frustrations can emerge when the next-nearest AF interaction, $J_2$, becomes active.
Theoretical studies on the $J_1$-$J_2$ honeycomb lattice predict the formation of a highly degenerate ground state, which can lead to a spin liquid state \cite{Okumura2010} or exotic multiple-$\bm{q}$ topological magnetic order\cite{Shimokawa2019}.
Recently particular focus is on Kitaev quantum spin liquid, which serves as a solvable theoretical model for a two-dimensional (2D) honeycomb layer with Ising interactions, as represented by ${\alpha}$-RuCl$_3$ and iridates\cite{Chaloupka2010, Jackeli2009, Williams2016}.
However, finding an ideal Kitaev compound is challenging due to small distortions from the honeycomb lattice and/or non-negligible interlayer interactions in actual compounds, which can lead to the formation of specific orders. 
Considerable efforts have been undertaken to investigate ideal 2D honeycomb lattice compounds.

\begin{figure*}[!t]
\centering
        \includegraphics[width=15cm]{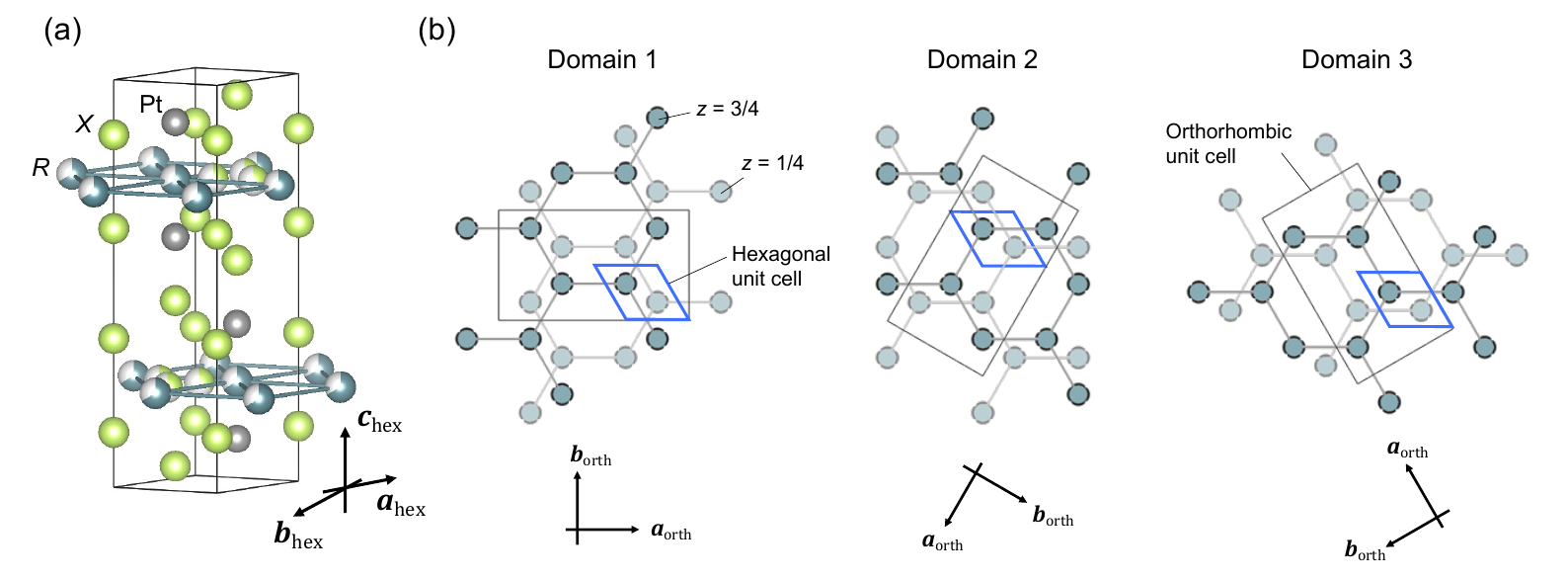}
    \caption{ Crystal structure of {\rpx} drawn by VESTA \cite{Momma2011}. (a) Average hexagonal structure. (b) Orthorhombic structure with three domains rotated by 120$^{\circ}$ each around $c_{\rm hex}$ ($c_{\rm orth}$). Pt and $X$ atoms are not shown.}
    \label{fig:crystal_structure}
\end{figure*}

In this context, {\rpx}($R$=rare earths, $X$=Ga, Al) could provide an interesting playground. 
{\rpx} crystallizes in the hexagonal Sc$_{0.6}$Fe$_2$Si$_{4.9}$-type structure with the space group $P6_{3}/mmc$ (no. 194, $D_{6}^{h4}$) \cite{Radzieowski2017,Radzieowski2020}, illustrated in Fig. \ref{fig:crystal_structure}(a).
One remarkable feature of this structure is that $R$ and $X$ atoms have a regular vacancies of 1/3 and 2/3, respectively.
These systematic vacancies result in the formation of a honeycomb network for the $R$ atoms \cite{Haga2020,Matsumoto2021,Kwei1996}.
The unit cell contains two well-separated honeycomb layers of $\sim$8 \AA \ along the $c$-axis.
Note that this stacking has a slight shift along [01$\bar{1}$0].
As there exists three equivalent [01$\bar{1}$0], there appears three domains in terms of stacking as displayed in Fig. \ref{fig:crystal_structure} (b).
Consequently, while the symmetry of the local unit cell is orthorhombic, an average structure with three domains results in a hexagonal symmetry.
To avoid confusion, subscripts ``hex" and ``orth" are used to represent averaged hexagonal and orthorhombic crystal systems, respectively.
The relationship between the hexagonal and orthorhombic cells is depicted in Fig. \ref{fig:crystal_structure} (b).
The presence of the three domains indicates intrinsic stacking disorder of this structure.
Together with the long separation between honeycomb layers, weak interlayer interaction is expected, in other words, {\rpx} appears likely a host of quasi-2D magnetic behavior.
Indeed, diffuse scattering along the $l$ direction around (1/3, 1/3, $l$)$_{\rm hex}$ reflections was confirmed in the X-ray diffraction experiments, indicating short range correlations along the $c$-axis\cite{Radzieowski2020,Matsumoto2021}.

Here we focus on the new 5$f$ variant {\upg}.
{\upg} also has a stacking disorder with lattice constants of $a_{\rm hex}~=~4.304$~\AA \ and $c_{\rm hex}~=~16.30$~\AA \cite{Matsumoto2021}.
Magnetization measurements reveal strong uniaxial nature of this compound with the easy axis along $c_{\rm hex}$ \cite{Matsumoto2021}.
Despite a lack of the structural long-range correlation along the $c$-axis, a bulk crystal exhibits a fairly sharp transition at {\tn}~=~26~K as observed in magnetization, specific heat, and electrical resistivity measurements.
This sharp transition implies a dominant in-plane and relatively weak interlayer exchange interactions, as pointed out by the previous study of bulk physical properties \cite{Matsumoto2021}.
To our best knowledge, an uranium compound with a 2D honeycomb structure is rarely known.
{\upg} may provide interesting ordering phenomena under frustration between 5$f$ electrons with strong spin-orbit interaction.
In order to get microscopic insights into magnetic order in {\upg}, resonant X-ray scattering (RXS) and neutron diffraction experiments were carried out on single crystalline samples.

\section{Experiment}
Single crystalline \upg \ was grown by the Ga-self flux method, of which details were reported in Ref. \onlinecite{Matsumoto2021}. 
Rod-like crystals with a typical dimension of 1 $\times$ 1 $\times$ 3 mm$^3$ ($\sim$5 mg) were used in both RXS and neutron diffraction experiments.
These crystals also display similar (1/3, 1/3, $l$)$_{\rm hex}$ diffuse scatterings, consistent with those reported in the literature \cite{Radzieowski2020,Matsumoto2021}.
The RXS experiments were performed at BL-11B of Photon Factory, using a two-circle diffractometer for soft X-ray  installed inside a high-vacuum chamber \cite{Nakao_2014}.
The typical vacuum level in the chamber was  $10^{-5}$ Pa throughout the measurements.
An incident X-ray energy was set to around 3.72~keV to fulfill a resonance condition of the U $M_4$ edge.
The two samples with different orientations were attached to Cu blocks by using silver paste, and the Cu blocks were mounted onto a cold finger of a liquid $^4$He flow cryostat with base temperature of about 6~K. 
A horizontal scattering geometry with the incident $\pi$-polarization, in which the electric field vector lying in the scattering plane, was employed.
The polarization of the scattered X-ray was analyzed using an Al (111) single crystal.

The neutron diffraction experiments were carried out on the thermal triple-axis spectrometer TAS-1 installed at the reactor hall of the research reactor JRR-3 of Japan Atomic Energy Agency in Tokai, Japan.
A vertical focusing PG monochromator and analyzer were used.
The spectrometer was fixed to the elastic condition in a triple-axis mode with a wavelength of 2.40~\AA.
The collimation set of open-80'-40'-80' was employed, and higher order contamination was removed by a PG filter placed before the sample.
The single crystalline sample was fixed to an Al plate to have a ($hhl$) horizontal scattering plane.
The sample was sealed in an Al can filled with a He exchange gas, and was attached to a cold finger of a closed-cycle refrigerator and cooled down to 3~K.

\section{Results}
\subsection{Resonant X-ray scattering}
\begin{figure}[!t]
\centering
        \includegraphics[width=8cm]{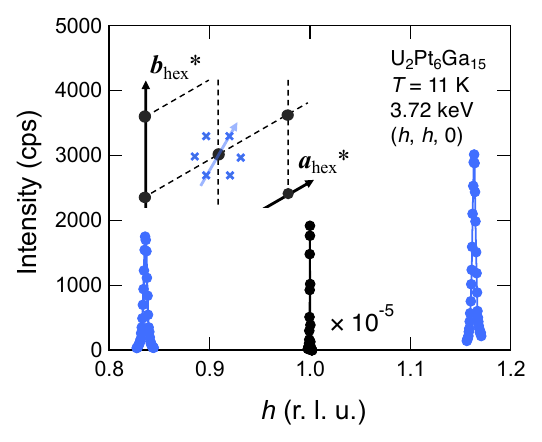}
    \caption{The line scan profile along [110]$_{\rm hex}$ of \upg \ measured under the resonant condition of the U M4 edge. The inset illustrates the observed diffraction pattern around the 110 fundamental reflection and the scan trajectory.}
    \label{fig:hh0_line}
\end{figure}

\begin{figure}[!t]
\centering
        \includegraphics[width=7.5cm]{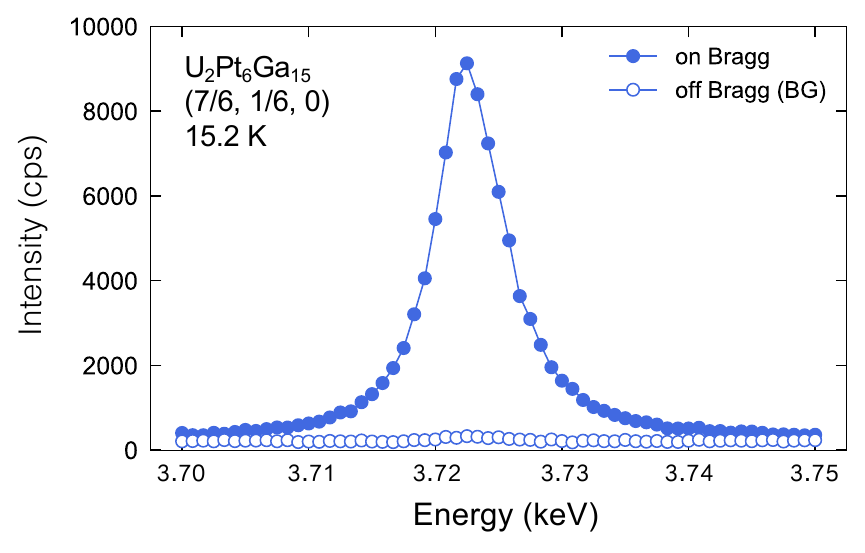}
    \caption{Energy dependence of the resonant X-ray magnetic scattering at $(7/6, 1/6, 0)_{\rm hex}$ (closed circles), shown with the fluorescence background (open circles).}
    \label{fig:Escan}
\end{figure}

\begin{figure}[t]
\centering
        \includegraphics[width=8.5cm]{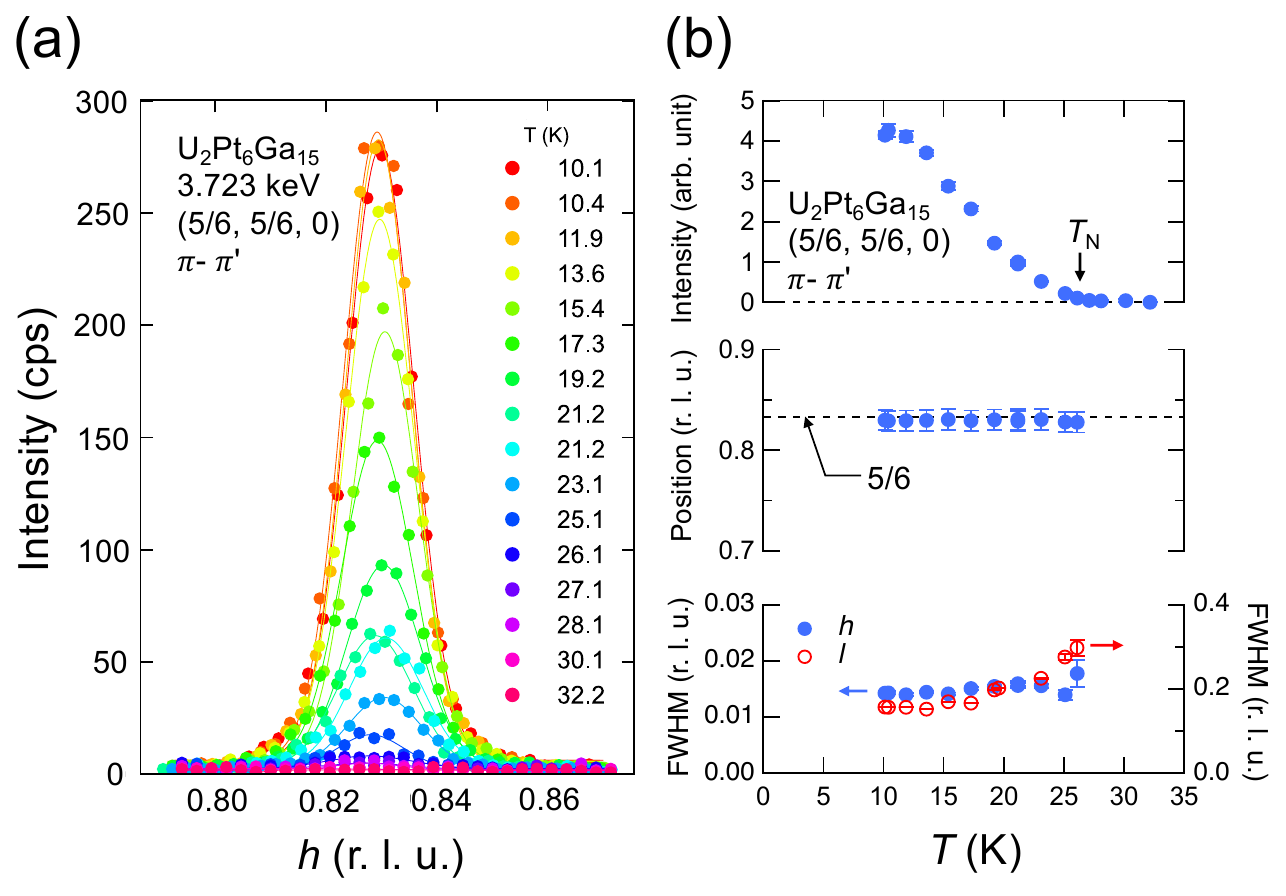}
    \caption{Temperature dependence of (a) peak profiles and (b) integrated intensity, peak positions and full width at half maximum (FWHM) of the resonant X-ray magnetic scattering at $(5/6, 5/6, 0)_{\rm hex}$.}
    \label{fig:T-dep_RXS}
\end{figure}

Superlattice reflections originating from magnetic ordering were searched by scans along the high symmetry reciprocal lattice axes, [100]$^*_{\rm hex}$, [001]$^*_{\rm hex}$, and [110]$^*_{\rm hex}$ at the resonant condition of the U $M_4$ absorption edge at 11~K.
Here, the experimental data are represented in the hexagonal notation due to the intrinsic difficulty in distinguishing individual orthorhombic domains in diffraction experiments.
Figure \ref{fig:hh0_line} shows a scan profile along [110]$_{\rm hex}$.
Satellite peaks were found at (5/6,~5/6,~0)$_{\rm hex}$ and (7/6,~7/6,~0)$_{\rm hex}$ besides a fundamental peak at (1,~1,~0)$_{\rm hex}$.
Additional scans confirmed that superlattice reflections form a six-fold symmetry around (1,~1,~0)$_{\rm hex}$ in the $a^{*}_{\rm hex}$-$b^{*}_{\rm hex}$ plane, reflecting the average hexagonal symmetry, as illustrated by the inset of the figure.
This result indicates that the propagation vector of the AF state can be described by $\bm{q} = (1/6, 1/6, 0)_{\rm hex}$.
Figure \ref{fig:Escan} shows the peak intensity of the typical satellite reflection (7/6,~1/6,~0)$_{\rm hex}$ as a function of incident X-ray energy.
The well-defined peak around 3.722~keV corresponds to the U $M_4$ absorption edge, indicating that the satellite originates from a resonant scattering involving 5$f$ electrons. 
The temperature variation of the satellite peaks is depicted in Fig.~\ref{fig:T-dep_RXS}.
Upon cooling, the RXS signal intensity gradually increases below $T_{\rm N}$, indicative of order parameter development.
This increase in intensity is not accompanied by a peak shift, as shown in the right panel in Fig. ~\ref{fig:T-dep_RXS} (b).
Namely, the ordering vector $\bm{q}$ remains commensurate with $\bm{q}=(1/6, 1/6, 0)_{\rm hex}$ across the entire temperature range below {\tn}.
The peak width remains constant along the $h$ direction throughout the temperature variation, while it slightly broadens with increasing temperature in the $l$ direction.

\begin{figure}[t]
\centering
        \includegraphics[width=8.5cm]{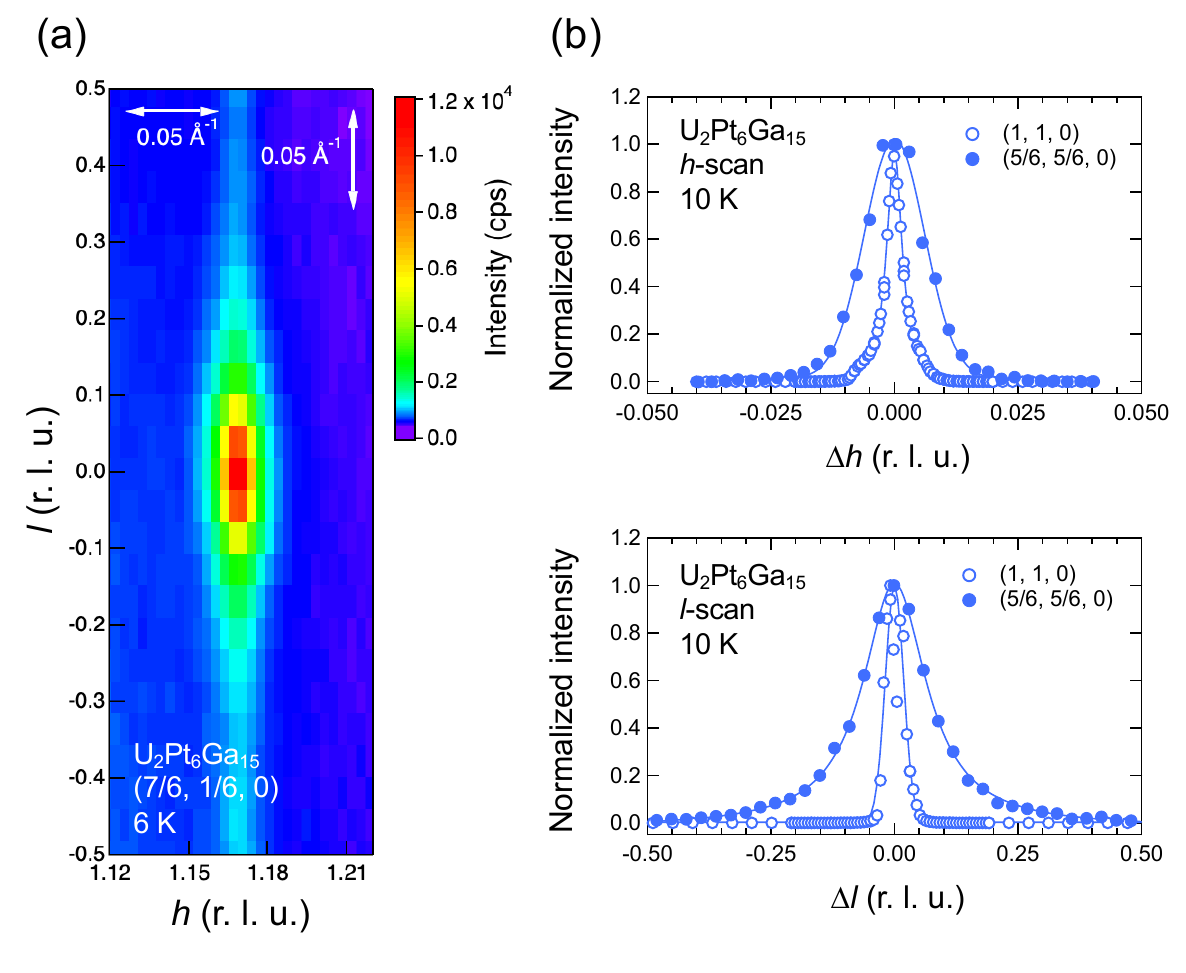}
    \caption{(a) A color map of the RXS peak of \upg \ in the $(h, 0, l)_{\rm hex}$ reciprocal lattice plane. (b) Line profiles of the RXS peak and the fundamental reflection peak along $(h, 0, 0)_{\rm hex}$ and $(0, 0, l)_{\rm hex}$ lines.}
    \label{fig:streak}
\end{figure}

The observed magnetic reflections have a highly anisotropic peak shape in reciprocal space as represented by the ($h,0,l$) intensity map for the (7/6, 1/6, 0)$_{\rm hex}$ reflection in Fig. \ref{fig:streak} (a). 
The peak has a streak along the $l$ direction at 6~K, indicating that short-range correlations remain along the $c_{\rm hex}$-axis well below {\tn}.
Figure~\ref{fig:streak} (b) displays the peak profiles along the [100]$^*_{\rm hex}$ and [001]$^*_{\rm hex}$ directions, together with those of the (1, 1, 0)$_{\rm hex}$ fundamental reflection.
The magnetic reflection is broader than the fundamental reflection for both directions.
Corresponding to the streak in the color map, the profile along the $l$-direction is substantially broadened with a Lorentzian form.  
The correlation lengths of the magnetic order, estimated from the full width at half maximum (FWHM), are approximately 70 unit cells (260 \AA) along the $a_{\rm hex}$-axis and 7 unit cells (100 \AA) along the $c_{\rm hex}$-axes.
Given the high resolution of the synchrotron X-rays, the negligible impact of its resolution on the observed peak width suggests that the width predominantly reflects the intrinsic nature of the sample, revealing the 2D nature of magnetic correlations in {\upg}.

\begin{figure}[t]
\centering
        \includegraphics[width=8.5cm]{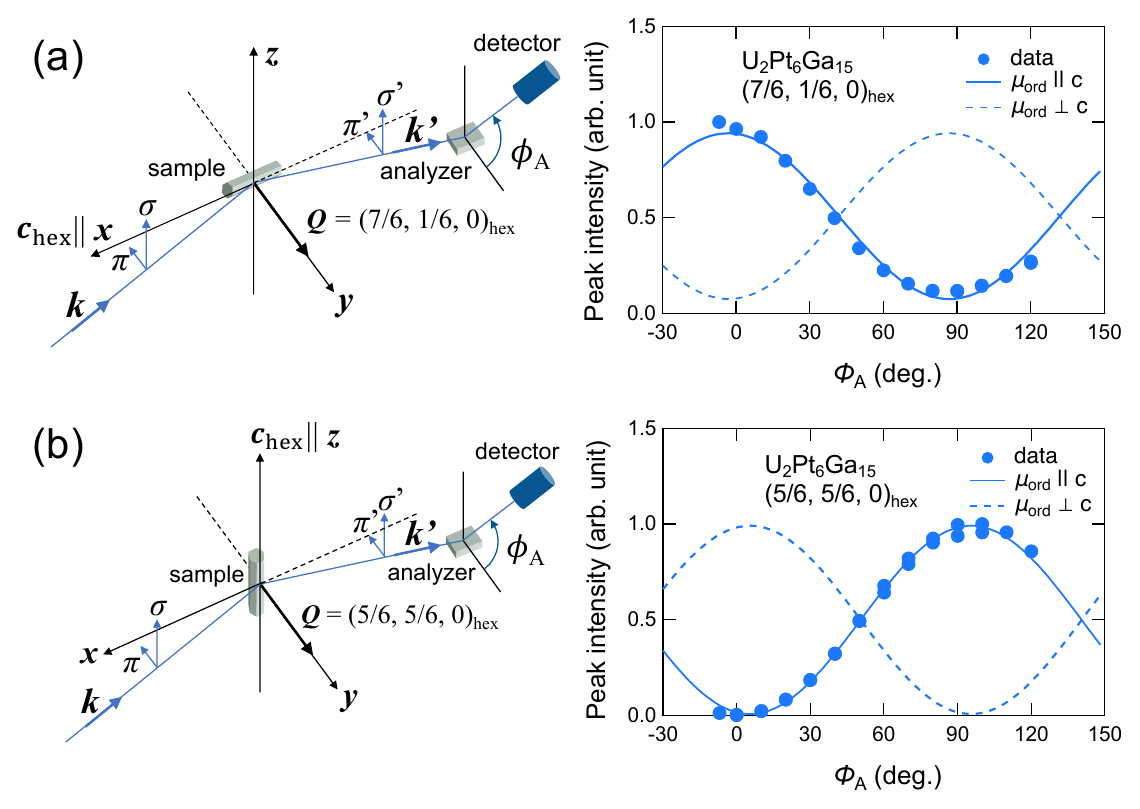}
    \caption{Polarization analyses of the RXS signal for the magnetic scatterings measured at 10 K. The schematic view of geometric configurations of the sample and optical system are also shown for two configurations with the $c_{\rm hex}$-axis (a) parallel and (b) perpendicular to the scattering plane. The vectors of $\bm{k}$, $\bm{k'}$, and $\bm{\kappa}$ are wave vectors of the incident and scattered X-ray, and the scattering vector, respectively.}
    \label{fig:pol}
\end{figure}

To investigate orientation of the ordered moment, the polarization of the scattered X-ray was analyzed in two configurations: $c_{\rm hex}$ (a) within and (b) normal to the horizontal scattering plane.
Figure~\ref{fig:pol} shows the RXS intensity as a function of the analyzer crystal angle, $\phi_{\rm A}$, for the two scattering geometries depicted in Fig.~\ref{fig:pol}(a) and (b).
Polarization components $\pi'$ and $\sigma'$ of the scattered X-ray were obtained for $\phi_{\rm A}$~=~90$^{\circ}$ and 0$^{\circ}$, respectively.
As shown in the figure, the two configurations give contrasting dependencies on $\phi_{\rm A}$.
In configuration (a), the maximum intensity was observed at $\phi_{\rm A}$=0$^{\circ}$, corresponding to $\sigma'$ polarization, while in configuration (b), it was observed at $\phi_{\rm A}$=90$^{\circ}$, corresponding to $\pi'$ polarization.

The observed polarization dependences in both configurations consistently demonstrate the magnetic moments parallel to the $c_{\rm hex}$ axis.
First, the strong RXS intensity in the $\pi$-$\sigma'$ process in configuration (a) indicates that a rotation of the polarization vector of the X-ray occurs. 
This cannot be caused by the Thomson's scattering process, and necessitates magnetic dipoles or even higher-rank multipoles. 
Strong intensity in neutron diffraction indicates magnetic dipole origin, as detailed in the following section.
The finite intensity in the $\pi$-$\pi'$ channel is a background signal which is not intrinsic to the magnetic order, as evidenced by the absence of peak structure in the rocking curve.

Next, $\phi_{\rm A}$ dependence in configuration (b) is discussed based on the orientation of the magnetic dipole moment relative to the polarization.
The RXS amplitude arising from magnetic dipole is given by $-i(\bm{\varepsilon}' \times \bm{\varepsilon}) \cdot \bm{u}$ \cite{Hannon1988}, where ${\bm{\varepsilon}}$ and $\bm{{\varepsilon}}'$ are polarization vectors of the incident and scattered X-ray, respectively, and $\bm{u}$ is a unit vector parallel to the magnetic moment.
The absence of detectable signal in the $\pi$-$\sigma'$ strongly suggests that the ordered moments are pointing to the $c_{\rm hex}$ axis.
Indeed, the Ising-like moments along to the $c_{\rm hex}$ axis, suggested by the bulk measurements\cite{Matsumoto2021}, well reproduce the observed $\phi_{\rm A}$ dependences as illustrated by the dotted lines in Figs.~\ref{fig:pol}(a) and (b).

\subsection{Neutron scattering}
The RXS experiment succeeded to determine the propagation vector $\bm{q}$ and the orientation of the ordered moment.
On the other hand, a long wavelength of X-ray for the U M$_4$ edge limits an accessible reciprocal space.
A complete magnetic structure, namely the spin arrangement within the unit cell, could not be determined owing to this restriction.
To determine complete spin structure, the neutron diffraction experiment was performed on a single crystal.

\begin{figure}[!t]
\centering
        \includegraphics[width=8.5cm]{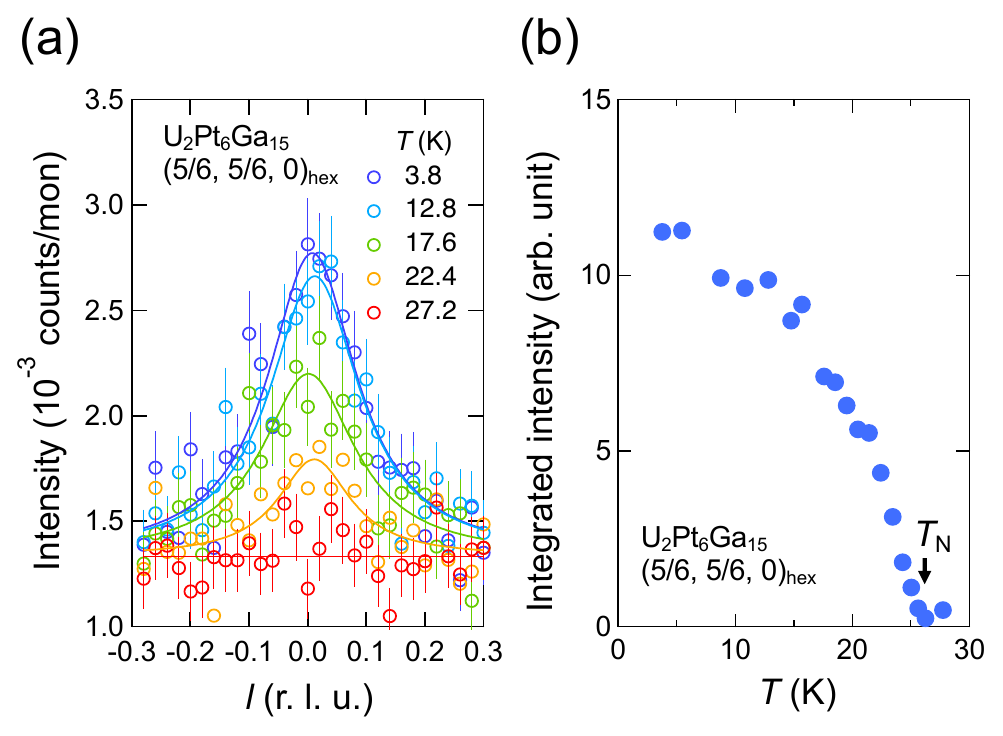}
    \caption{Temperature dependence of (a) the peak profile and (b) integrated intensity of the neutron magnetic reflection at $\bm{q} = (5/6, 5/6, 0)_{\rm hex}$ of \upg.}
    \label{fig:T-dep_NS}
\end{figure}

Figure~\ref{fig:T-dep_NS}(a) shows peak profiles of the magnetic Bragg reflection along $l$ through (5/6,~5/6,~0).
The magnetic reflection with $\bm{q} = (1/6, 1/6, 0)_{\rm hex}$ which develops below {\tn} was confirmed in the neutron scattering as well.
The peak profile is reproduced by a broad Lorentzian.
The integrated intensity as a function of temperature in Fig.~\ref{fig:T-dep_NS}(b) reveals gradual evolution of the magnetic peak below {\tn}~=~26~K.
This increase is slightly steeper in neutron than RXS, which may reflect different penetration length of two probes.
On the other hand, {\tn} and the broad nature of the magnetic peak are consistent between two probes, indicating that the observed short range correlation stems not from surface effect, but from bulk.

\begin{figure}[!t]
\centering
        \includegraphics[width=7cm]{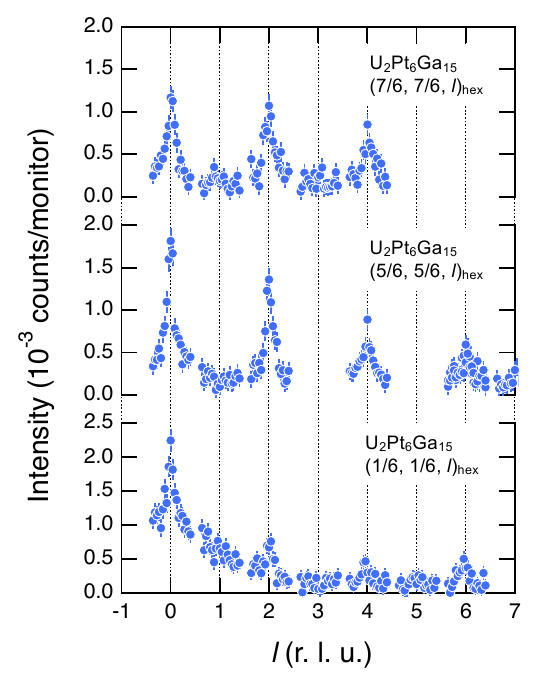}
    \caption{Line profiles along [001] in the reciprocal lattice space measured at the base temperature, 3.8 K. The top, middle, and bottom panels show (1/6, 1/6, $l$), (5/6, 5/6, $l$), and (7/6, 7/6, $l$) lines, respectively.}
    \label{fig:linescans}
\end{figure}

To determine the magnetic structure, magnetic peaks in wide $\bm{Q}$-space were measured at the base temperature.
Figure \ref{fig:linescans} shows line scan profiles along $l$ for (1/6, 1/6, $l$)$_{\rm hex}$, (5/6, 5/6, $l$)$_{\rm hex}$, and (7/6, 7/6, $l$)$_{\rm hex}$ at 3~K.
In all three scans, magnetic peak intensities decrease with increasing $l$, and is more pronounced for (1/6, 1/6, $l$)$_{\rm hex}$.
This feature is consistent with the Ising-type magnetic structure as magnetic neutron diffraction intensity is proportional to the square of the magnetic moment perpendicular to the scattering vector $\bm{Q}$.

One remarkable feature is a systematic extinction; intense magnetic peaks were observed only at $l$ = even for all three scans.
The systematic absence of the magnetic peak at $l$=odd could reflect characteristic structure along the $c$-axis.
U ions in the unit cell can be regarded to form a double-layered structure along $c$, though it does not fulfill precise translational symmetry with (0,~0,~1/2)$_{\rm hex}$ and have a slight displacement on stacking. 
The observed systematic absence with $l$=odd suggests that the two layers in the unit cell have a ferromagnetic coupling.
Details of a magnetic structure model will be discussed in the next section.

\section{Discussion}
First, a possible magnetic structure of {\upg} is discussed in this section.
Typically, a straightforward quantitative analysis of magnetic Bragg intensities from a neutron diffraction measurement gives possible magnetic structure models.
There exists two difficult factors, however, in the present case.
One is the elongation of the magnetic Bragg peaks along $l$, that makes it hard to deduce reliable absolute intensities by conventional scans. 
Another difficulties arise from domains in terms of stacking displacement of the U layer along $c_{\rm hex}$ ($c_{\rm orth}$).
As mentioned before, the honeycomb layers do not stack up perfectly along $c_{\rm hex}$ ($c_{\rm orth}$), but are shifted along $b_{\rm orth}$, resulting in the formation of three domains rotated by 120$^{\circ}$ with each other (see Fig. \ref{fig:crystal_structure} (b)).
Because the present neutron experiment covered only first quadrant of the 2D scattering plane in the reciprocal space, effects of domain could not be taken into account.

\begin{figure*}[t]
\centering
        \includegraphics[width=15cm]{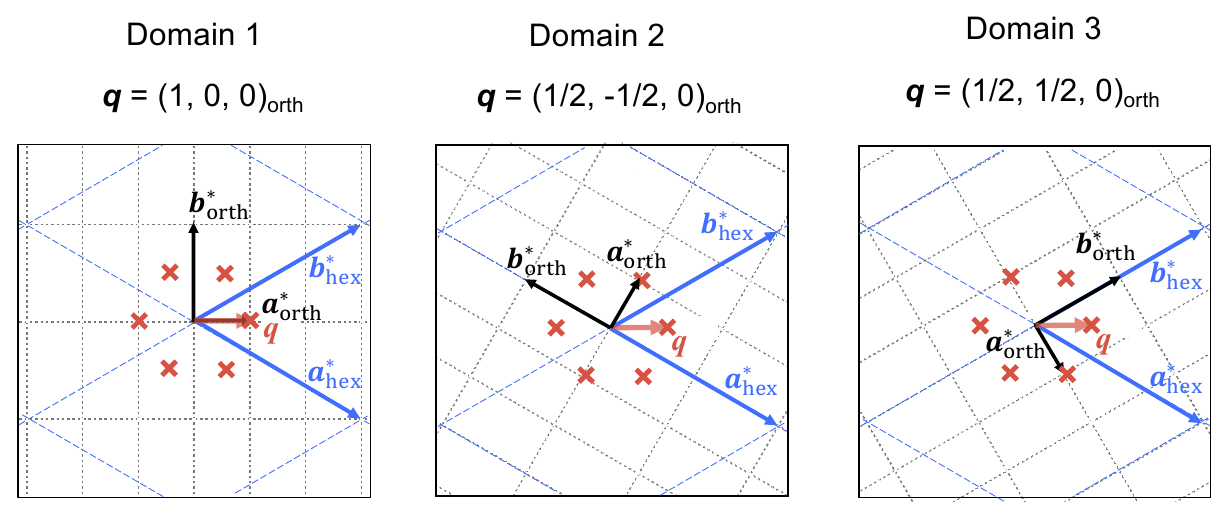}
    \caption{Schematic of the relationship between the local orthorhombic lattice system and the average hexagonal lattice system for \upg. The cross symbols indicate the positions of the observed magnetic scattering. The closed circles represent the squared structure factor calculated for the orthorhombic structure.}
    \label{fig:Qspace}
\end{figure*}

\begin{figure}[h]
\centering
        \includegraphics[width=8cm]{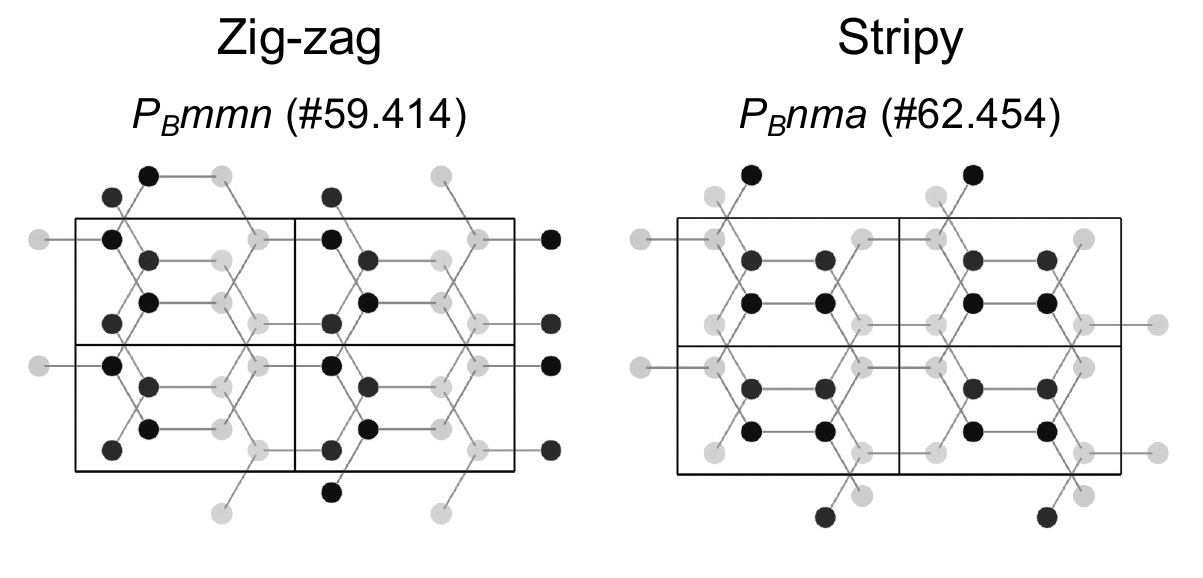}
    \caption{Magnetic structure candidates for \upg \ suggested by the RXS and neutron diffraction experiments. The black circles and light grey circles indicate down spins and up spins, respectively. Four ($2 \times 2$) magnetic unit cells are shown for each.}
    \label{fig:structures}
\end{figure}

To overcome these intrinsic difficulties, we utilize a group theoretical approach. 
Whereas the domain structure prevents quantitative analysis, relationship between the propagation vector and the stacking offset could be determined.
Hereafter, the orthorhombic notation is employed in the following discussion to handle the stacking offset properly.
The present RXS and neutron diffraction experiments revealed following characteristics of the magnetic order of \upg:
\begin{enumerate}[(i)]
 \item The propagation vector of the magnetic order is $\bm{q} = (1/6, 1/6, 0)_{\rm hex}$.
 \item The ordered magnetic moments orient parallel to $c_{\rm hex}$ ($c_{\rm orth}$).
 \item The neighboring honeycomb layers in the unit cell couple ferromagnetically. 
\end{enumerate}
Owing to a choice on coupling between the stacking displacement and the magnetic propagation, 
there exists three possibilities to describe $\bm{q} = (1/6, 1/6, 0)_{\rm hex}$ in the orthorhombic notation, $\bm{q} = (1, 0, 0)_{\rm orth}$, $(1/2, -1/2, 0)_{\rm orth}$, and $(1/2, 1/2, 0)_{\rm orth}$, as displayed in Fig.~\ref{fig:Qspace}.
By applying the group theory for the result (i), 16 magnetic space groups are allowed, and the additional condition (ii) excludes half of the candidates\cite{Perez-Mato2015}.
The condition (iii) narrows down to two models both described by the propagation vectors $\bm{q} = (1, 0, 0)_{\rm orth}$, the zig-zag structure $P_{B}mmn$ (\#59.414) and the stripy structure $P_{B}nma$ (\#62.454), as displayed in Fig.\ref{fig:structures}.

Note that the two structures give identical intensity ratio for the observed magnetic reflections in (1/6, 1/6, $l$)$_{\rm hex}$, (5/6, 5/6, $l$)$_{\rm hex}$, and (7/6, 7/6, $l$)$_{\rm hex}$.
On the other hand, there appears a clear contrast for another high symmetry point, such as $(1/2, 1/2, 0)_{\rm hex}$, which corresponds to $(3, 0, 0)_{\rm orth}$.
A magnetic peak is not expected at $(1/2, 1/2, 0)_{\rm hex}$ for the zig-zag structure, whereas the stripy structure gives more than 3 times stronger intensity for $(1/2, 1/2, 0)_{\rm hex}$ than that for $(1/6, 1/6, 0)_{\rm hex}$.
Absence of a magnetic peak at $(1/2, 1/2, 0)_{\rm hex}$ in both neutron and RXS experiments indicates that the zig-zag structure is favorable as the magnetic structure in the ground state of {\upg}. 
The observed and calculated neutron diffraction intensities are summarized in Table~\ref{tab:int_ratio}.

\begin{table}[b]
\caption{Observed and calculated values of intensity ratios of magnetic reflections at (1/6, 1/6, 0)$_{\rm hex}$ and (1/2, 1/2, 0)$_{\rm hex}$.}
\renewcommand{\arraystretch}{2}
\begin{tabular*}{7cm}{@{\extracolsep{\fill}}llll}
\hline
\multirow{2}{*}{}                                             & \multicolumn{2}{c}{$I_{\rm calc}$} & \multirow{2}{*}{$I_{\rm obs}$} \\ \cline{2-3}
                                                              & Zig-zag          & Stripy          &                                \\ \hline
$I_{\frac{1}{2} \frac{1}{2} 0}/I_{\frac{1}{6} \frac{1}{6} 0}$ & 0.00             & 3.46            & 0.00(1)                        \\ \hline
\end{tabular*}
\label{tab:int_ratio}
\end{table}

Having determined the spin configuration, we now discuss the results in terms of the honeycomb lattice layers.
The zig-zag order in honeycomb layered systems has been found in several transition metal oxides such as ZnMnO$_3$ \cite{Haraguchi2019}, Na$_2$IrO$_3$ \cite{Liu2011}, and Na$_3$Co$_2$SbO$_6$ \cite{Wong2016}.
In these cases, ferromagnetic (FM) first-nearest neighbors ($J_1$) and AF second-nearest neighbors interactions ($J_2$) are considered to be essential.
The AF $J_2$ comparative to the FM $J_1$ brings frustration to a honeycomb network, preventing system from ordering in the N\'{e}el-type AF state.
Such a situation in an Ising spin system was theoretically analyzed within classical spin models\cite{Kudo1976}, pointing out that an introduction of a finite AF interaction between third-nearest neighbors, $J_3$, stabilizes the zig-zag order out of another degenerating phase with a doubled magnetic unit cell. 
Thus, the zig-zag type magnetic structure in the ground state in {\upg} can be basically understood within a classical Ising spin system, that is characterized by 2D nature with frustration.
The negative in-plane Weiss temperature, $\Theta_W{\perp} = -204$ K may be a consequence of the considerable AF interactions of $J_2$ and $J_3$,
although the crystalline electric field interaction should be taken into account.

It is worth mentioning the magnetic correlations between honeycomb layers and the effect of the stacking disorder in {\upg}.
The magnetic ordering itself is robust against the disorder.
This is demonstrated by the sharp transition at {\tn} in the bulk properties\cite{Matsumoto2021} and the consistency of {\tn} across samples in our RXS and neutron scattering results.
The key factor of the robustness may be the 2D nature in the magnetic interaction, namely, the order is primarily driven by the intralayer magnetic interactions.
This assumption finds support in the theoretical model that indicates the zig-zag order does not necessitate interlayer interactions\cite{Kudo1976}.
As suggested by the positive Weiss temperature along the $c$-axis ($\Theta_{\parallel} = +58$ K) \cite{Matsumoto2021}, an interlayer coupling could be FM.
Although it is not considered significant, the FM interlayer interaction can play a supportive role when the system favors the zig-zag structure as this structure accommodates the highest number of interlayer FM spin pairs, which maximizes energy gains in the interlayer coupling.
To elucidate overall spin exchange interactions in details, inelastic neutron scattering experiments to investigate spin excitation are in progress.

\section{Conclusion}
Magnetic order in a honeycomb layered antiferromagnet {\upg} was investigated using resonant X-ray and neutron scatterings on single crystalline samples.
Magnetic Bragg reflections described by a propagation vector of $\bm{q} = (1/6, 1/6, 0)_{\rm hex}$, which is $\bm{q} = (1, 0, 0)_{\rm orth}$ in the orthorhombic system, were observed in the both RXS and neutron scattering experiments.
The observed magnetic reflections revealed anisotropic magnetic correlation lengths with relatively short interlayer correlation reflecting the characteristic crystal structure with a systematic stacking disorder. 
The RXS experiment provided evidence of U 5$f$ electron ordering, as strong resonant scattering at the U $M_4$ absorption edge was detected.
The polarization dependence of the RXS signal explicitly indicates the ordered moments perpendicular to the honeycomb layer, consistent with the Ising-type anisotropy in the bulk magnetization.
The neutron scattering experiments, followed by group theoretical analysis, successfully determined that the spin structure in the ground state is of the zig-zag type.
The zig-zag structure and the quasi-two-dimensional correlation length suggest competing in-plane spin exchange interactions, involving both FM $J_1$, and AF $J_2$ and $J_3$ interactions, along with relatively weak interlayer coupling in this compound.

\begin{acknowledgments}
We acknowledge valuable discussions with Andrew D. Christianson. 
We would like to thank to Yoshinori Kitajima for assistance in using BL-11B of Photon Factory.
This work was supported by JSPS KAKENHI Grants Nos. JP23H04867, JP23H04871, JP21K14644, JP20H01864, JP21H01027, JP21H04987, JP23H01132, and JP23K03332. 
The synchrotron radiation experiments were performed at Photon Factory with the approval of Photon Factory Program Advisory Committee (Proposal No. 2020G072). 
\end{acknowledgments}

\bibliography{upg_ver03.bib}
\end{document}